%% file: reusing_is2023.tex
\documentclass{INTERSPEECH2023}
\usepackage{subcaption}
\usepackage{multirow} 

\interspeechcameraready

\title{Beyond Universal Transformer: block reusing with adaptor in Transformer for automatic speech recognition}
\name{Haoyu Tang$^1$, Zhaoyi Liu$^2$, Chang Zeng$^3$, Xinfeng Li $^4$}
\address{
  $^1$Department of Electronic Systems,
NTNU, Trondheim, Norway\\
  $^2$Computer Science, KU Leuven,  Leuven, Belgium $^3$National Institute of Informatics, Tokyo, Japan\\
  $^4$Electrical Engineering, Zhejiang University, Hangzhou, China 
  } 
\email{Charlie\_Tang\_1992@outlook.com, zhaoyi.liu@student.kuleuven.be, zengchang@nii.ac.jp, xinfengli@zju.edu.cn}

\begin{document}

\maketitle
\begin{abstract}
Transformer-based models have recently made significant achievements in the application of end-to-end (E2E) automatic speech recognition (ASR). It is possible to deploy the E2E ASR system on smart devices with the help of Transformer-based models. While these models 
still have the disadvantage of requiring a large number of model parameters. To overcome the drawback of universal Transformer models for the application of ASR on edge devices, we propose a solution that can reuse the block in Transformer models for the occasion of the small footprint ASR system, 
which meets the objective of accommodating resource limitations without compromising recognition accuracy. Specifically, we design a novel block-reusing strategy for speech Transformer (BRST) to enhance the effectiveness of parameters and propose an adapter module (ADM) that can produce a compact and adaptable model with only a few additional trainable parameters accompanying each reusing block. We conducted an experiment with the proposed method on the public AISHELL-1 corpus, and the results show that 
the proposed approach achieves the character  error rate (CER) of 9.3\%/6.63\% with only 7.6M/8.3M parameters without and with the ADM, respectively. In addition, we also make a deeper analysis to show the effect of ADM in the general block-reusing method.
\end{abstract}
\noindent\textbf{Index Terms}: speech recognition, Transformer, adapter module, layer-reusing

\input{section/introduction}
\input{section/related}
\input{section/method}
\input{section/exp}
\input{section/conclusion}

\bibliographystyle{IEEEtran}
\bibliography{mybib}

\end{document}

%% file: section/introduction.tex
\section{Introduction}
There is significant interest in developing automatic speech recognition (ASR) systems on smart devices to meet privacy, security, and network capacity limits. The advancements in E2E ASR systems indicate that such systems are now strong candidates for such deployments. Instead of independently building acoustic, language, and pronunciation models in traditional hybrid ASR systems, the E2E ASR systems integrate these components into a single sequence-to-sequence (seq2seq) model. Thanks to Transformer's \cite{vaswani2017attention} superior performance in processing sequence-related tasks \cite{raffel2020exploring}, Transformer-based models have been widely adopted in various state-of-the-art (SOTA) ASR systems and brought a recent breakthrough in E2E ASR \cite{zhang2020pushing}.

The advancements described in \cite{raffel2020exploring, gulati2020conformer} provide evidence that a large network is essential for obtaining SOTA performance \cite{ng2021pushing, guo2021recent}. However, the modeling ability of these approaches depends on a large number of parameters. The model with a large number of parameters makes it infeasible for deploying it on devices with limitations in memory and storage.



Currently, the optimization of E2E ASR system for smart devices has received greater attention, including model parallelization \cite{shao2020pychain}, knowledge distillation \cite{yang2022knowledge}, and low-rank factorization \cite{winata2020lightweight}, etc. Although these efforts have made considerable headway, the limitations of both memory and computing resources still make it challenging to widely deploy an ASR system on smart devices.

In this paper, we propose an extremely low footprint E2E ASR system with the block-reusing strategy inspired by \cite{dehghani2018universal, lan2019albert} for Transformer-based models to more effectively balance the storage and the precision of speech recognition. Specifically, compared with using multiple attention-based blocks in the universal Transformer based-model \cite{dong2018speech}, we adopt a block-reusing strategy, which means the encoder and the decoder in the Transformer possess only one block but pass through it multiple times when in forwarding propagation of both training and inference stages. 
This strategy is equivalent to treating each repetition as an independent block in the original baseline, which means using one block with multiple repeats to replace multiple blocks. It could dramatically decrease the model size. This proposed model is named block reusing (BR). But our later block-level analysis shows that this repeating technology same within \cite{dehghani2018universal, lan2019albert} will drop in local optimization points. To jump out of the local optimization, then multiple ADMs are inserted into the module to rectify the performance degradation for BR. This rectification is named as BRA method in this paper. In the later similarity analysis, an interesting founding is reported by us. Pure block reusing method in \cite{dehghani2018universal, lan2019albert} will push the reusing block deteriorating into an almost linear process, which is main reason for performance degradation. And the ADMs could dramatically enhance the non-linearity of the model.

We report extensive experiments on the public AISHELL-1 benchmark \cite{bu2017aishell}. Experimental results show that block-reusing with the ADMs improves parameter efficiency without sacrificing model accuracy. Our proposed LRST method achieves the word error rate (WER) of 9.3\%/6.63\% with only 7.6M/8.3M parameters without and with the ADM, respectively.

The rest of this paper is organized as follows. Related works are discussed in Section ~\ref{sec:Related}. Our proposed method is described in Section~\ref{sec:Method}. Experimental setup, result and discussion appear in Section~\ref{sec:Experiments}. Finally, Section~\ref{sec:Conclusion} provides our conclusions.

%% file: section/related.tex
\section{Related Works} \label{sec:Related}
\subsection{CTC-ATT Speech Transformer}
\begin{figure}
    \centering
    \includegraphics[width=\linewidth]{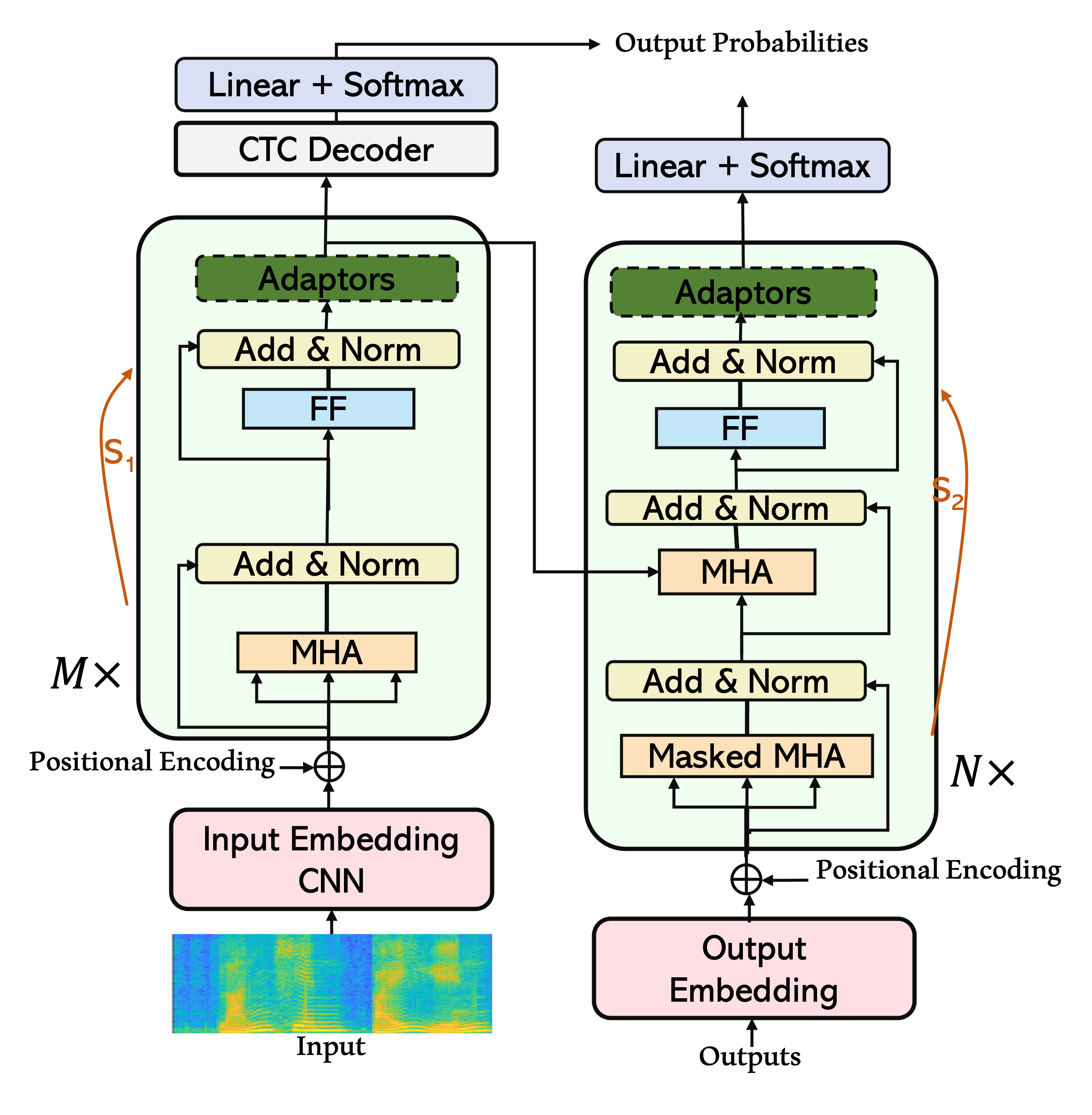}
    \caption{CTC-ATT transformer with LR/LRA}
    \label{fig:model}
\end{figure}
The CTC-ATT Transformer-based \cite{nakatani2019improving} block-reusing mechanism could be described in Fig.\ref{fig:model}. The encoder, containing $M$ blocks, maps the sequence of input speech feature $X=\left\{\mathbf{x}_{t} \mid t=1, \cdots, T\right\}$ to a sequence of embedding $H=\left\{\mathbf{s}_{l} \mid l=1, \cdots, L\right\}$. Rather than a vanilla Transformer with a standard single encoder-attention-decoder structure, there are two decoders separately mapping embedding to the character.

The Connectionist Temporal Classification (CTC) decoder will map embedding $\textbf{H}$ into CTC topology sequence $C=\left\{c_{l} \in \mathcal{U} \mid l=1, \cdots, L\right\}$ with a set of distinct character, $\mathcal{U}$. It should be noted this character set not only contains a linguistic unit but also a "blank symbol" unit which explicitly denotes the letter boundary to handle the repetition of character symbols \cite{graves2006connectionist}. Meanwhile, the original attention decoder could simultaneously or later on generate one element at each time consuming on the embedding $H$, this auto-regressive decoder continues complete a sequence of token $Y=\left\{\mathbf{y}_{s} \in \mathcal{U} \mid s=1, \cdots, S\right\}$. These two decoders actually share the same character set $\mathcal{U}$. As a seq2seq model, letters boundary naturally exist within the sequence but the attention decoder should still have the "blank symbol" in its character set for later decoding.

The multi-objective function of the original CTC-ATT Transformer is implemented in \cite{watanabe2017hybrid} and later on migrate into speech-Transformer for ASR \cite{nakatani2019improving}. Within the training stage of the CTC-ATT Transformer, rather than a data-driven attention training force model to learning alignment ignoring the monotonic speech-text natural properties, the CTC training plays a role as an auxiliary task to speed up the process of estimating the desired alignment. The optimized function of CTC-ATT is a simply logarithmic linear combination of CTC decoder and attention decoder objective:
\begin{equation}
    \mathcal{L}_{\mathrm{CTC-ATT}}=\lambda \log p_{\mathrm{ctc}}(C \mid X)+(1-\lambda) \log p_{\mathrm{att}}^{*}(C \mid X) \label{equa:loss_combine}
\end{equation}

There are several decoding methods for the CTC-ATT speech transformer. With these decoding algorithms, the decoders can generate token sequence $\textbf{C}$ consuming embedding $\textbf{H}$. In the attention decoding algorithm, the attention decoder could generate one element at each time. And this auto-regressive decoder continues a sequence of the token. In CTC greedy search decoding, just greedy searching is the best path in the CTC decoder's mapping. Usually, beam search has a precisely global path search.  Also, CTC prefix beam search merge sequences have the same non-blank token sub-sequence in each time step. The attention rescore decoding is designed for combining the n-best beam logarithmic probability with attention decoder logarithmic probability by text in the beam. The shared character set $\mathcal{U}$ in two decoders could easily linearly combine probability which is the reason attention still keeps the "blank symbol".

In vanilla CTC-ATT Transformer, the encoder, and attention decoders respectively contain $M$ transformer encoder blocks and $N$ transformer decoder blocks. And each block contains exclusive parameters which make the model large. It should be noted parameters get involved in forwarding propagation just once.

In the BR strategy, within the encoder, each block actually shares the same set of parameters across blocks to make the block have high parameter efficiency. Similar strategies have been reported in \cite{dehghani2018universal}. Based on simple block reusing,  \cite{lan2019albert} also explore only block reusing, it shows that if only one share attention, the performance will drop a little, but only decrease the number of parameters by around one-third. And then \cite{gao2021extremely} explore the possibility migrate its application field from language to speech. It should be noted that \cite{zhao2021non} also points out that the transformer encoder blocks have high similarity. 

%% file: section/method.tex
\section{Method} \label{sec:Method}
In this section, the details of block-reusing will be introduced, to obtain a super low footprint E2E ASR system. How the adaptors in the encoder and attention decoder alleviate deterioration of WER is discussed likewise.

\subsection{Block-reusing Strategy in Transformer} \label{subsec:blockreuse}
In our design, this high parameter efficiency reusing also could be reviewed as a block reusing since actually there is only a block but repeat forward $M$ times. Basically, the total storage and training parameters in the encoder can be theoretically reduced into $1/M$ of the one without weight sharing. The blocks in the attention decoder also could share parameters across blocks and reduce the parameters to $1/N$. 

\subsection{Block-reusing Strategy with adapter module}
The function of the original CTC-ATT transformer encoder could be stated as:
\begin{equation}
\label{equa:encoder_compose1}
    H = f(X) = f_{M}(f_{M-1}(f_{M-2}(......f_{1}(X))) 
\end{equation}
In the equation, the $f_{m}$ represents the $m$-th block of the encoder, so the $f$ could represent the whole encoder.

As declared in the \cite{zhao2021non}, there is enormous similarity within the blocks in the encoder. And \cite{houlsby2019parameter} explores the adaptors in language transformer finetune. Inspired by the block similarity and adaptor, block function could be decomposed as a combination of public function and a unique function, and described as follows:
\begin{equation}
    f_m(x) = f_{m'}(f_{0}(x)) \label{equa:block_compose1}
\end{equation}
Within the equation, the $f_{0}$ is the public function for all blocks, and $f_{m'}$ is the ADM function of block $m$.

If the $f_{m'}$ supposed as identification function, which means the all the block function is same, the function \ref{equa:block_compose1} can be view as:
\begin{equation}
    f_m(x) = f_{0}(x) \label{equa:block_compose2}
\end{equation}
Then replace the $M$ blocks with one block but $S_1$ is repeated. The equation \ref{equa:encoder_compose1} degenerate:
\begin{equation}
    H = f(X) = \underbrace{f_{0}(f_{0}(f_{0}(......f_{0}(X)))}_{S_1\ times} \label{equa:encoder_compose2}
\end{equation}
It is clear that equation \ref{equa:encoder_compose2} basically is BR described in subsection \ref{subsec:blockreuse}. Nevertheless, it is based on the assumption all the blocks are the same. Without this assumption, the blocks' function are still viewed as a similar function, the the equation \ref{equa:encoder_compose1}\ would be combined with equation \ref{equa:block_compose1}, then the whole encoder equation \ref{equa:encoder_compose2}  mapping is rewrite as:
\begin{equation}
        H = f(X) = f_{M'}(f_{0}( f_{(M-1)'}(f_{0}(f_{1'}(f_{0}(x))))
        \label{equa:encoder_compose3}
\end{equation}
Based on the equation \ref{equa:encoder_compose3}, the ADM will be inserted as function $f_{M'}$. As mentioned, the $f_{0}$ is the reused encoder block. So this function actually is the mathematical representation of the block reusing adaptor module (BRA). Meanwhile, ADM function could be proposed in the attention decoder as well with a reused encoder block with repeated $S_2$ times.

%% file: section/exp.tex
\section{Experiments} \label{sec:Experiments}
\subsection{Experiments Setup}
There is speech recognition tasks for the evaluation proposed method, AISHELL-1 \cite{bu2017aishell}.

The acoustics feature used in all experiments is 80-dimensional log-Mel filter bank (fbank) energies computed with 25ms windows width and 10ms windows shift. The characters set $\mathcal{U}$ in both datasets are Chinese characters simply generated by statistics all Chinese characters in its corresponding train text file. The baseline CTC-ATT model is composed of a 12 blocks encoder and 6 blocks decoder. Each block in encoder and attention and decoder outsize all are 256, with 4 heads in multi-head attention. And its corresponding BR(A) are showed in the table \ref{table:Model_hyper_parameters}. The BR/BRA models only contain 1 encoder block but are repeated 12 times and 1 decoder block is repeated 6 times shown in the table \ref{table:Model_hyper_parameters}.\\
\begin{table}[t!]
    \centering
    \caption{Model hyper-parameters}
    \begin{tabular}{|c|c|c|c|c|} \hline
         \textbf{Model} & \textbf{$M$} & \textbf{$N$} & \textbf{$S_{1}$} & \textbf{$S_2$} \\ \hline
         baseline & 12 & 6 & 1 & 1 \\ \hline
         BR(A) & 1 & 1 & 12 & 6 \\ \hline 
    \end{tabular}
    \label{table:Model_hyper_parameters}
\end{table}
In the training phase, the $\lambda$ in linear loss combine equation \ref{equa:loss_combine} is set as 0.3. The optimizer of the model is a warm-up Adam optimizer with a learning rate of 0.002 and 25,000 warm-up steps. The original waveform is augmented with speed-perturb with ratio 0.9 and 1.1 during training phase. Meanwhile, the train feature fbank is pre-processed by specaugment \cite{park2019specaugment, park2020specaugment}. There are two frequency-dimension masks and two time-dimension masks. The maximum mask in time and frequency respectively are 50 and 10 bins. 

\subsection{AISHELL Result}
\input{section/AISHELL_wer_table.tex}
After challenge, since the test set is not public, so we move to a public Mandarin dataset for further development and explore. Initially, we can implemented a BR model its size still just 1/3 of baseline showing in table \ref{table: CER_AISELL_test}, but CER increase much more from baseline if compared with last dataset. At this point, the ADM method is introduced. First of all, we found that the CER increasing at CTC greedy and CTC prefix is much more than others. Compared with these two path, encoder $\xrightarrow{}$ CTC decoder, and encoder $\rightarrow{}$ attention decoder, two path actually share a same encoder. Since CTC decoder is just a small linear output later, there are much more bias in encoder actually, but attention decoder rectify some of them. 

At next step, the ADM is implemented as a simple linear+ReLu and inserted into the encoder. This experiment named as "BRA-E" in table \ref{table: CER_AISELL_test}. Of course this insertion slightly gain the number of parameters. It is clear that it basically relieved the CER increasing in all decoding method since it push encoder to have a better feature extraction. Especially in attention rescore CER decrease around 26.3\% percent CER from BR experiment. Then the linear+ReLu ADM is inserted not only encoder and also attention decoder. In additional, there are two more experiments, "BRA-ED" and "BRA-D". Literally, the experiments with encoder and decoder both inserted ADM named as "BRA-ED" and with only decoder inserted ADM named as "BRA-D". In the "BRA-ED" have all most same experiment result with "BRA-E" which means the ADM in decoder don't have enough help. Within the "BRA-D", it is clear the ADM can not help if it only rectify at shallow block.

For verifying the gradient clip tuning, we also reduced gradient clip from 5 to 3 as well. Unfortunately, gradient clip experiment in "BRA-E" and "BRA-ED" don't evident improvement in this noiseless dataset. Inspired by \cite{gao2021extremely, lan2019albert}, the $S_1$ also is increase from 12 to 18 to explore the further improvement. Unfortunately, its CER also almost keeps same. But compared with the similar setup Exp6 ($M=N=1$) in \cite{gao2021extremely}, our proposed method "BRA-E" almost keeps the similar CER but just 65.4\% the number of parameters in model. 

\subsection{Similarity Analysis}
For analysis the mapping processing of ADM, we extract $m$th block's embedding in baseline and $m$th repeated reusing block embedding in BR/BRA to compare their similarity with AISHELL testset. The similarity metrics is linear centered kernel alignment (CKA) introduced in \cite{kornblith2019similarity} for compare representation in deep learning model. 
\begin{figure}[t!]
    \centering
    \begin{subfigure}[b]{0.45\textwidth}
         \centering
         \includegraphics[width=\textwidth]{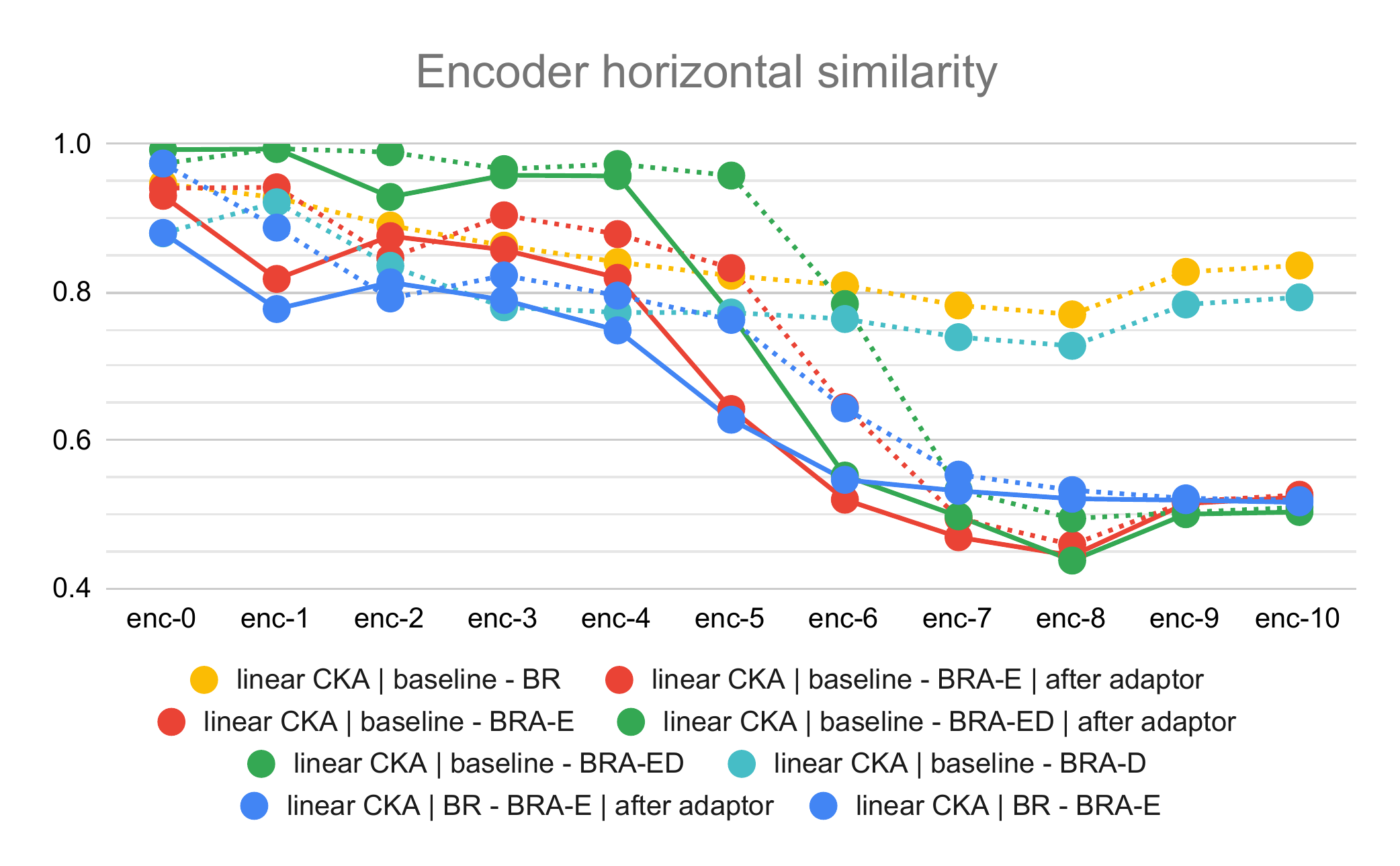}
         \caption{Encoder} \label{subfig:CKA_horizontal_encoder}
    \end{subfigure}
    \hfill
    \begin{subfigure}[b]{0.45\textwidth}
         \centering
         \includegraphics[width=\textwidth]{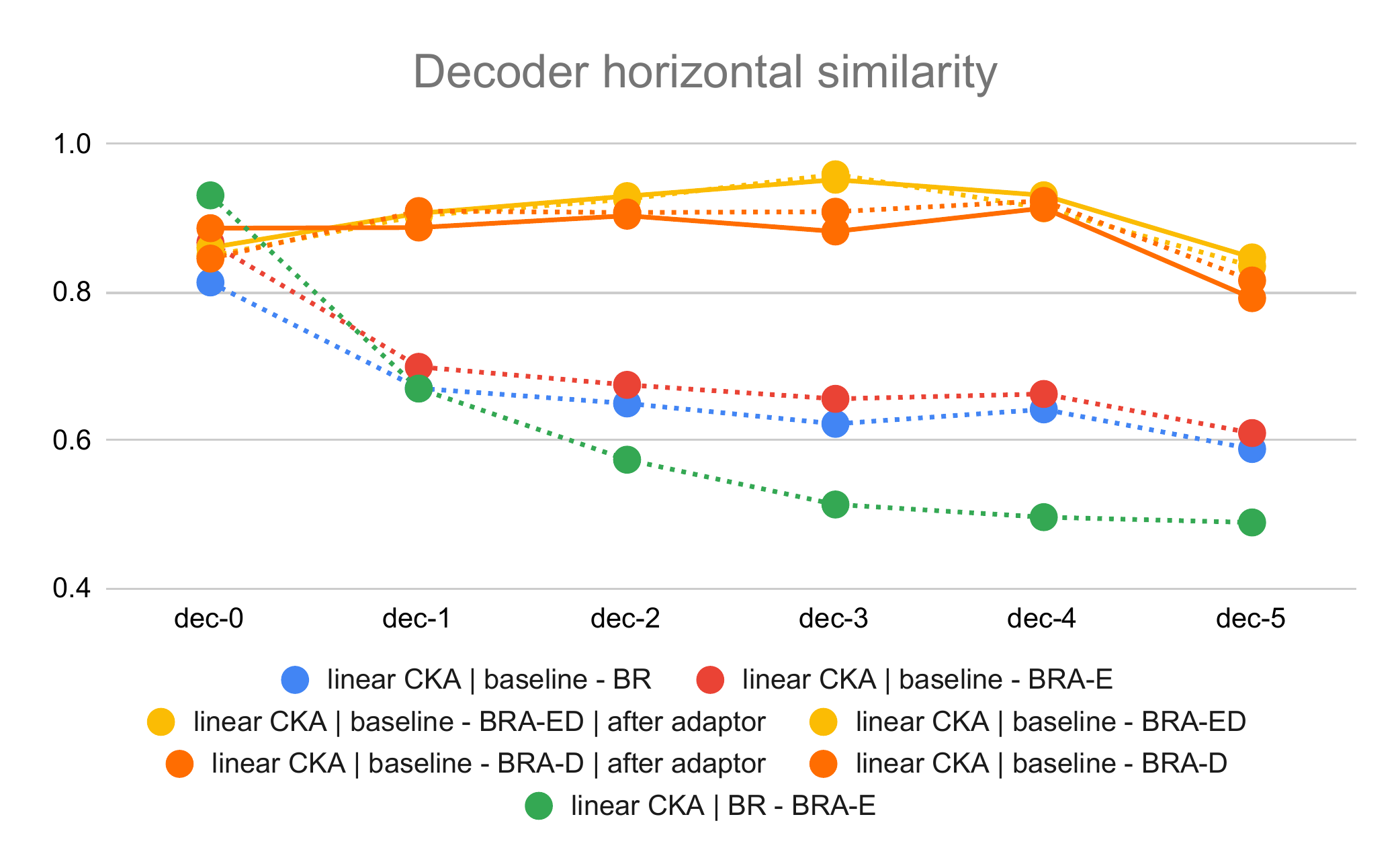}
         \caption{Decoder} \label{subfig:CKA_horizontal_decoder}
    \end{subfigure}
    \caption{Models horizontal similarity in AISHELL test}
    \label{fig:block_horizontal_CKA}
\end{figure}
Except the each proposed method models' embedding are compared with baseline model, BR and BRA-E models' embedding are contrasted. The compared metrics are showed in the figure \ref{fig:block_horizontal_CKA}. The horizontal axes "enc/dec-i"show the the $m$th similarity as mention in encoder or decoder. The different lines with colors show model comparison pair, more specifically, the dotted lines simple show similarity comparison between from $m$th block's embedding with $m$th repeating's embedding.

More specifically the solid lines tailed with "after ADM" means similarity comparison of the $i$th block's embedding in baseline with $i$th ADM embedding. In the similarity analysis, define a AMD push-away number, as the difference between a solid line with its corresponding dotted line in each block. A large number of push-away number shows the ADMs indeed push repeating block's representation away from its original distributing. The ADMs indeed enhance the diversity of embedding for repeated blocks. Also it could be explained as the $f_{m'}$ equation \ref{equa:block_compose1} have a strong non-linear representation ability.

In encoder, the baseline - BR/BRA-D lines almost follows a same trend, slowly decreasing with model getting deeper, also reported in \cite{lan2019albert}. Within new introduced ADM, BRA-E(D) transformer block' similarity lines also drops in deeper blocks with a higher slope.

Intuitively, if a lite model compressed from a large model, it will be generally think as its represent should as close as to original represent for each compressed part. This hypothesis could be demonstrate in the Figure \ref{fig:block_horizontal_CKA} as, the solid lines always have a much more higher value with its correspond dotted lines. But the Figures \ref{subfig:CKA_horizontal_encoder} shows that the BRA-E(D) models have a much bigger ADMs push-away number as the hypothesis's expectation, but actually the dotted line have a much more strong similarity with baseline. ADMs do enhance the diversity for each repeating's representation, but it do not make distribution close to baseline. However, even BRA-E(D) have a mush have different distribution from baseline with BR model, these model achieve a lower CER in the end. In the other hand, it could be reviewed as the ADMs push models out of local optimization points.

In decoder, the baseline - BR/BRA-E lines drops slowly with deeper block as well, similar with corresponding BR/BRA-D in encoder. The funny things is, the AMD push-away number in BRA-(E)D actually is quite smell, which means the ADMs in decoder cannot enhance the diversity. That is might the reason there is improvement from BRA-E $\rightarrow{}$ BRA-D
 and BR $\rightarrow{}$ BRA-D experiments.

And this BR - BRA-E difference could be cross-valid in the green "linear CKA | BR- BRAE" for both encoder and decoder as well.

\begin{figure}[t!]
    \centering
    \begin{subfigure}[b]{0.45\textwidth}
         \centering
         \includegraphics[width=\textwidth]{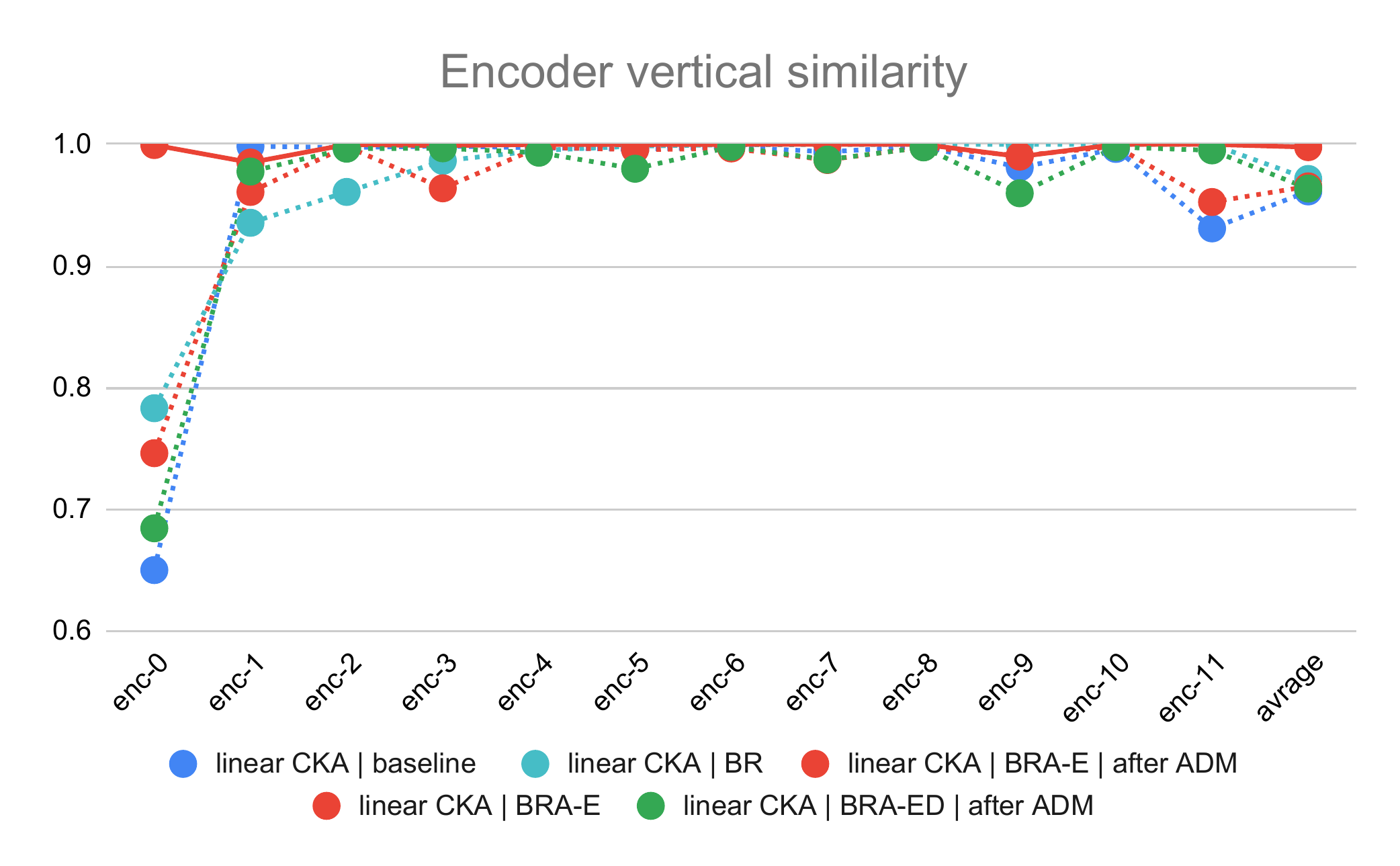}
         \caption{Encoder} \label{subfig:CKA_vertical_encoder}
    \end{subfigure}
    \hfill
    \begin{subfigure}[b]{0.45\textwidth}
         \centering
         \includegraphics[width=\textwidth]{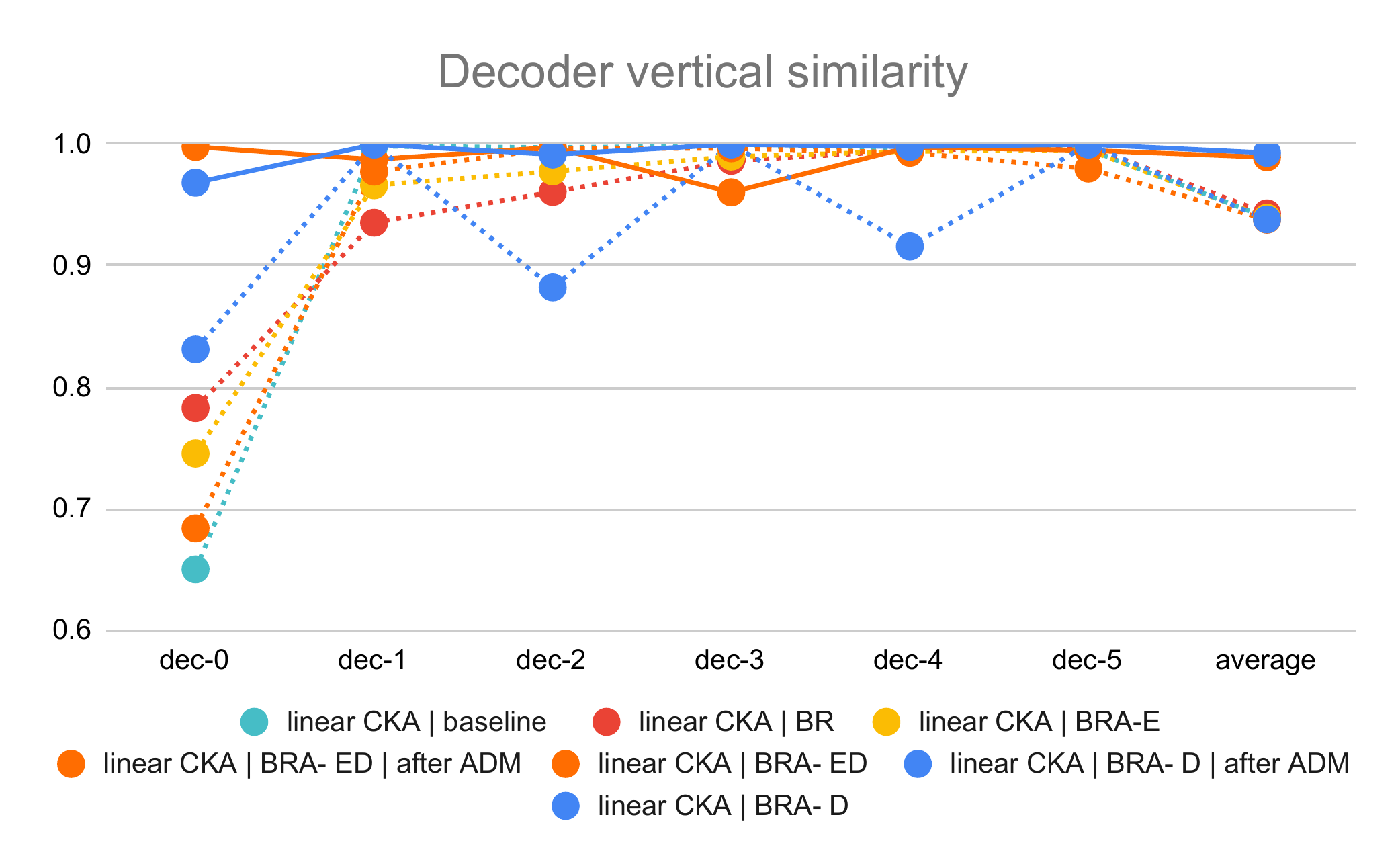}
         \caption{Decoder} \label{subfig:CKA_vertical_decoder}
    \end{subfigure}
    \caption{Models vertical similarity in AISHELL test}
    \label{fig:block_vertical_CKA}
\end{figure}

Besides of horizontal similarity, we also want to know how each repeating and ADMs models. So the input and output of each block are compared shown in the figure \ref{fig:block_vertical_CKA} named with model vertical similarity analysis with same metrics. The transformer block process similarity is shown in the figure as dotted lines, and ADMs process similarity is shown as solid lines similar to figure \ref{fig:block_horizontal_CKA}. It should be pointed out that if there is a strong similarity between input and output of a block, which means input and output have strong linear similarity, and this block could be viewed as an approximate linear transformation.

In the encoder, it is clear that all the first blocks except BRA-E(D) transformer block have a strong non-linear. And then all lines keep a near 1 horizontal line, which means the actually non-linear process is mainly finished in the first block in all models. The reason the BRA-E(D) transformer first transformer block keeps strong linearity is the first AMD plays the role for it. It also could be a very important idea of why the BR model degrades so much from checkpoint, simple change blocks into one block repeating will makes this block the trade-off between in each repeating role. It will make this block deteriorate and lose non-linear modeling called \textbf{linear deterioration}. This discovery could explain the model degradation in the \cite{dehghani2018universal, lan2019albert, gao2021extremely} as well.

In the decoder, similar result within encoder, BR still shows linear deterioration, and just replace BRA-E(D) with BRA-(E)D. The biggest difference is the AMDs' in BRA-D show alternate linear and non-linear.

In the end, all ADMs ever in encoder and decoder show a stronger non-linearity than transformer block repeating.

\begin{table}[t!]
    \centering
    \caption{BRA initialized from BR}
    $^*$ means initialized from BR \\
    \begin{tabular}{|c|cccc|} \hline
        \multirow{2}{*}{Exp} & \multicolumn{4}{c|}{CER (\%)} \\ \cline{2-5} 
        & \multicolumn{1}{c|}{CG} & \multicolumn{1}{c|}{CP} & \multicolumn{1}{c|}{ATT} & ATT-RE \\ \hline
        BRA-E$^*$& \multicolumn{1}{c|}{31.36} & \multicolumn{1}{c|}{31.1}& \multicolumn{1}{c|}{26.9} & 33.01 \\ \hline
        BRA-ED$^*$& \multicolumn{1}{c|}{32.01} & \multicolumn{1}{c|}{31.69} & \multicolumn{1}{c|}{33.82} & 27.63 \\ \hline
    \end{tabular}
    \label{table: BRA_finetune_from_BR}
\end{table}

For valid similarity repeating blocks from BR and BRA, we train a BRA-E and BRA-ED model initialized with BR checkpoint showing in table \ref{table: BRA_finetune_from_BR}. It clearly shows there is much degrade in BR initialized BRA-E and BRA-ED models. This degrade cross verifies that actually the BRA's representation is much different from BR's representation.

%% file: section/AISHELL_wer_table.tex
\begin{table}[] \footnotesize
\centering
\caption{The WER of LR/LRA model in AISHELL test}
$^*$ The CG, CP, ATT, ATT-RE respectively means the decoding method: CTC greedy, CTC prefix, attention, attention resore \\
$^{**}$ the EXP6 result in the Table 2 in \cite{gao2021extremely}, similar setup with all BR/BRA experiments
\begin{tabular}{|c|c|cccc|}
\hline
\multirow{2}{*}{Model}                                & \multirow{2}{*}{Params (Mb)} & \multicolumn{4}{c|}{CER $^{*}$(\%)}                                                                       \\ \cline{3-6} 
                                                      &                           & \multicolumn{1}{c|}{CG}         & \multicolumn{1}{c|}{CP}         & \multicolumn{1}{c|}{ATT} & ATT-RE     \\ \hline
baseline                                              & 29                        & \multicolumn{1}{c|}{5.92}       & \multicolumn{1}{c|}{5.91}       & \multicolumn{1}{c|}{5.69}      & 5.30 \\ \hline
BR                                                    & 7.75                      & \multicolumn{1}{c|}{11.75}      & \multicolumn{1}{c|}{11.7}       & \multicolumn{1}{c|}{8.07}      & 9.30 \\ \hline \hline
BRA-E                                                 & 8.5                       & \multicolumn{1}{c|}{\textbf{8.38}}       & \multicolumn{1}{c|}{\textbf{8.38}}       & \multicolumn{1}{c|}{\textbf{6.63}}      & 6.85 \\ \hline
\multicolumn{1}{|c|}{+ $S_1=18$ } & 8.875 & \multicolumn{1}{c|}{8.40}       & \multicolumn{1}{c|}{8.39}       & \multicolumn{1}{c|}{6.58}          & \multicolumn{1}{c|}{\textbf{6.78}} \\ \hline
BRA-ED                                                & 9                         & \multicolumn{1}{c|}{8.55}       & \multicolumn{1}{c|}{8.56}       & \multicolumn{1}{c|}{6.97}      & 7.07                  \\ \hline
BRA-D                                                 & 8.25                      & \multicolumn{1}{c|}{10.77}      & \multicolumn{1}{c|}{10.77}      & \multicolumn{1}{c|}{7.43}      & 8.55                  \\ \hline \hline
\multicolumn{1}{|c|}{\cite{gao2021extremely}$^{**}$}  & 13                        & \multicolumn{1}{c|}{\textbackslash}           & \multicolumn{1}{c|}{\textbackslash}           & \multicolumn{1}{c|}{6.52}      & \multicolumn{1}{c|}{\textbackslash} \\ \hline
\end{tabular} 
\label{table: CER_AISELL_test}
\end{table}

%% file: section/conclusion.tex
\section{Conclusion} \label{sec:Conclusion}
In this paper, we design a new block reusing method for transformer in E2E ASR system for memory and storage constrained device and enhance it with an adaptor module. It could achieve an extremely high parameters efficiency and decrease the number of parameters to one-third of the original one. This method also could prevent the number of model parameters from growing with the depth growing. The simple block reusing transformer block obtains an acceptable WER in AISHELL1. Then the adaptor module is imported to deliver more diversity to the model's representation which dramatically boosts the model performance. The model is evaluated in the AISHELL1 benchmark, obtaining a 6.63\% CER but only 8.5\%M parameters in the model. For understanding the degradation of WER in simple block reusing, the vertical and horizontal similarity is conducted. It not only shows that pure block reusing would trap the model in local optimization since simple block repeating lacks diversity, but also describe that the pure block reusing will lose non-linear representation as well.

In further work, we would like to focus on the parameter reusing method from block-wise to layer level, deep analysis of the layer in transformer block to realize a higher parameter efficiency, lower CER, and finally evaluate the methods in more scenarios.